%%%%%%%%%%%%%%%%%%%%%%%%%%%%%%%%%%%%%%%%%%%%%%%%%%%%%%%%%%%%%%%%%%%%%%
%                                                                    %
%     Space-Time Quantization and Nonlocal Field Theory              %
%   ---Relativistic Second Quantization of Matrix Model---           %
%                                                                    %
%                     by Sho Tanaka                                  %
%                     TeX with PHYZZX macro                          %
%%%%%%%%%%%%%%%%%%%%%%%%%%%%%%%%%%%%%%%%%%%%%%%%%%%%%%%%%%%%%%%%%%%%%%
\input phyzzx.tex
\hoffset=10mm
\vsize=26cm
\titlepage 
\title{{\bf Space-Time Quantization and Nonlocal Field Theory} \break
 ---Relativistic Second Quantization of Matrix Model---}
\author{Sho TANAKA\footnote{*}{Em.~Professor of Kyoto University and 
Associate Member of Atomic Energy Research Institute, Nihon University.
E-mail: stanaka@yukawa.kyoto-u.ac.jp}}
\address{Kurodani 33-4, Sakyo-ku, Kyoto 606-8331, Japan}
\abstract{We propose relativistic second quantization of matrix model 
of D particles in a general framework of nonlocal field theory based 
on Snyder-Yang's 
quantized space-time. Second-quantized nonlocal field is in general 
noncommutative with quantized space-time, but conjectured to become 
commutative with light cone time $X^+$. This conjecture enables us to 
find second-quantized Hamiltonian  of D particle system and Heisenberg's 
equation of motion of second-quantized {\bf D} field in close contact 
with Hamiltonian given in matrix model. We propose Hamilton's principle 
of Lorentz-invariant action of {\bf D} field  and investigate what 
conditions or approximations are needed to reproduce the above Heisenberg's 
equation given in light cone time. Both noncommutativities appearing 
in position coordinates of D particles in matrix model and in quantized 
space-time will be eventually unified through second quantization of 
matrix model.}

\endpage
\chapter{Introduction}
 In the previous paper\ref\Tanaka(referred to as I, hereafter) the present 
\def\Tanaka{S.~Tanaka, Nuovo Cim.{\bf 114B} (1999) 49; hep-th/9808064.}
author has reexamined the original idea of nonlocal field and clarified 
the necessity of introducing the concept of noncommutative space-time 
for the comprehensive description of extended objects. In fact, in the 
original bilocal field theories\ref\markov,\Ref\Y\yukawa nonlocal field 
$U$ was characterized by noncommutativity with space-time coordinates 
$X_\mu$, that is,
$$
                   [U, X_\mu] \neq 0,
\eqn\eqA
$$
discarding the restriction that field quantities are simply functions 
of space-time coordinates, but retaining the classical nature of the 
latter space-time coordinates. Consequently, as was shown by 
Yukawa\refmark{\Y}, 
if one chooses the representation basis vector, $ \vert x_\mu 
\rangle $, with simultaneous eigenvalues  $x_\mu$  of all the space-time 
coordinate components $X_\mu$ , let us call it the space-time coordinate 
representation hereafter, one  immediately arrives at  bilocal field 
$\langle  x'_\mu \vert U \vert x''_\mu \rangle  \equiv U(x'_\mu, x''_\mu)$,
 i.e., a simple two-point function of space-time coordinates, which is 
clearly insufficient to describe extended objects in general. 
\def\yukawa{H.~Yukawa, Prog.~Theor.~Phys.~{\bf 2} (1947), 209.}
\def\markov{M.~A.~Markov, Jour.~Phys.~{\bf 2} (1940), 453.}

This consideration naturally leads us to the idea of the noncommutative 
space-time, which directly invalidates the use of the above space-time 
coordinate representation and enables us to remove the limitation on 
the concept of nonlocal fields mentioned above. It should be further 
noticed that under the idea of noncommutative space-time, the original 
assumption $\eqA$  itself becomes a natural consequence of the idea, 
because nonlocal field $U$, even if it is naively a function of the 
noncommutative space-time, can never be commutative with space-time 
coordinates. 

As was pointed out in I, the idea of noncommutative space-time was early 
proposed by Snyder\Ref\S\snyder and Yang\Ref\Ya\yang in terms of ``quantized 
space-time'' in a Lorentz-covariant way. 
\def\snyder{H.~S.~Snyder, Phys.~Rev.~{\bf 71} (1947), 38; {\bf 72} (1947),
68.}
\def\yang{C.~N.~Yang, Phys.Rev. {\bf 72} (1947), 874; Proc.~of International 
Conf.~on Elementary  Particles, 1965 Kyoto, pp 322-323.}
According to their idea, we tried in I to describe both nonlocal field 
$U$ and 
noncommutative space-time quantities as operators working on the common 
linear space of a certain infinite series of functions of Snyder-Yang's 
parameters and to represent them in terms of noncommutative 
infinite-dimensional matrices.

\def\wi{E.~Witten, Nucl.~Phys.~{\bf B460} (1995), 335; hep-th/9510135.}
\def\bfss{T.~Banks, W.~Fischler, S.~H.~Shenker and L.~Susskind, Phys.~
Rev.~{\bf D55} (1997), 5112; hep-th/9610043.}
\def\sw{N.~Seiberg and E.~Witten, ''String Theory and Noncommutative 
Geometry,'' hep-th/9908142. In this article, one can find a comprehensive 
list of papers on noncommutative geometry.}
\def\min{D.~Minic, "Towards Covariant Matrix Theory," hep-th/0009131.}

In the present paper, we wish to extend this idea and apply it to the 
recent matrix model. This model is expected to describe quantum-mechanical 
many-body system of D-0 brane or D-particle, a typical extended object 
which is nowadays widely believed to be fundamental constituent of 
superstrings.\Ref\Bf\bfss\ As is well known, the idea of matrix model 
of position coordinates of D-branes actually motivated the concept of 
noncommutative space-time \Ref\winum\wi\ and the recent remarkable studies 
on noncommutative geometry.\ref\sw\ In the present paper, however, we 
take a viewpoint slightly different from the latter approach, and propose 
Lorentz-invariant second-quantized field theory of matrix model on the 
basis of noncommutative and quantized space-time of Snyder-Yang's type 
from the beginning. It turns out that second quantization of matrix 
model as quantum mechanical system of many D-particles is crucial in the 
Lorentz-covariant formulation of the latter model, the importance of which 
has been emphasized so far by several authors.\refmark{\Bf} \ref\min

The plan of the present paper is as follows. In section 2, first we briefly 
recapitulate the quantized space-time algebra ${\cal R}$ originally proposed 
by Snyder\refmark{\S} and Yang.\refmark{\Ya} We introduce Hilbert space 
I, on which quantized space-time quantities ${\cal R}$ work as operators. 
In section 3, nonlocal field $U$ is defined as an operator working on 
Hilbert space I on an equal footing with $\cal R$. The light cone 
representation bases are introduced in Hilbert space I, where both of 
nonlocal field $U$ and light cone time operator $X^+$ are assumed to 
become diagonal. This framework enables us to construct second-quantized 
theory of nonlocal field ${\bf U}$  according to Hamiltonian formalism 
in Hilbert space II as in the usual quantum field theory. In section 4,
 the idea is applied to the infinite number of D particle system described 
by matrix model, whose Hamiltonian is now given in terms of second-quantized 
${\bf D}$ field in light cone representation bases. In section 5, we 
propose Lorentz invariant action of ${\bf D}$ field and investigate what 
conditions or approximations are needed to reproduce Heisenberg's equation 
of motion given in section 4. The final section is devoted to conclusions 
and discussions.

\chapter{Recapitulation of Quantized Space-time and Quasi-local 
Representation}

As was emphasized in I, Snyder\refmark{\S} first challenged the idea 
of quantized space-time, by expressing space-time quantities by linear 
differential operators on the five-dimensional background space. Later 
on, Yang\refmark{\Ya} developed the idea by extending the background 
space to the six-dimensional space $(\xi_0,\xi_1,\xi_2,\xi_3,\eta_,\tau)$,
 where both space-time coordinate $X_\mu$ and momentum (or translation) 
$P_\mu$ operators were  defined as linear differential operators on the 
space:
$$
      X_i = i(\xi_i\partder{}{\eta}-\eta\partder{}{\xi_i}),
     \eqn\eqB
$$
$$
     X_0 = i(\xi_0\partder{}{\eta}+\eta\partder{}{\xi_0})
     \eqn\eqC
$$
and
$$
    P_i = -i(\xi_i\partder{}{\tau} -\tau \partder{}{\xi_i}),
    \eqn\eqD
$$
$$
    P_0 = -i(\xi_0\partder{}{\tau}+\tau\partder{}{\xi_0}).
    \eqn\eqE
$$ 
We call hereafter the space $(\xi_0,\xi_1,\xi_2,\xi_3,\eta_,\tau)$ and 
its dimensional extension in general, Snyder-Yang's parameter space.

 Now let us recapitulate, for the later convenience, the typical commutation 
 relations between the above quantities characteristic of the noncommutative 
 space-time:   
$$
    [X_i,X_j]=-iL_{ij}
   \eqn\eqF       
$$
$$
    [X_i,X_0]=iM_i,
   \eqn\eqG
$$
and 
$$
    [P_i,P_j]=-iL_{ij}
    \eqn\eqH
$$
$$
     [P_i,P_0]=iM_i,
     \eqn\eqI
$$
with $ L_{ij} $ and $ M_i $ defined by
$$
       L_{ij}=i(\xi_i\partder{}\xi_j - \xi_j\partder{} \xi_i)
       \eqn\eqJ
$$
and
$$
       M_i =i(\xi_0\partder{}\xi_i + \xi_i\partder{}\xi_0).
       \eqn\eqK
$$
 $L_{ij}$ and $M_i$ as a whole constitute six Lorentz generators, satisfying 
 the well-known Lorentz algebra, under which $ X_{i,0}$ and $ P_{i,0} 
 $ are transformed as Lorentz four-vectors. 
Furthermore, 
$$
     [X_i,P_j] =i \delta_{ij} N,
        \eqn\eqL
$$
$$
         [X_0,P_0] =-i N,
\eqn\eqM
$$
$$
        [X_i,P_0]=[P_i,X_0]=0.
\eqn\eqN
$$
Here one finds the operator $N$ 
$$
        N=i(\eta \partder{}\tau - \tau \partder{}\eta)
\eqn\eqO 
$$
is clearly Lorentz-scalar and plays the role of the usual quantum mechanical 
Planck constant $h$, and further it causes the reciprocal transformation 
between the coordinates and momenta as
$$
        [X_i,N] =-iP_i, \cdots
\eqn\eqP
$$   

It is important to note that the fifteen operators introduced above
$$
          {\cal R}_{15}\equiv (X_\mu,P_\mu,L_{\mu \nu},N)
\eqn\eqQ
$$
with $\mu=(i,0)$ and $L_{\mu \nu} \equiv (L_{ij},M_i)$,  constitute as 
a whole a Lie algebra, which is interpreted as a maximal set of Killing 
vector fields realized on a pseudosphere in six-dimensional Minkowski-space; 
$ -{\xi_0}^2 +{\xi_1}^2+{\xi_2}^2+{\xi_3}^2+{\eta^2}+{\tau}^2=constant. 
$ 

The original Snyder's algebra is now understood as one missing $\tau$ 
in the above Yang's model. Four-momenta $P_\mu $ are there defined as 
$P_i={\xi_i / \eta}, P_0={\xi_0 / \eta}$, while space-time coordinate 
operators being the same as \eqB\ and \eqC. They show, therefore, different 
commutation relations from Yang's ones; for instance, $[X_i,P_j] =-iP_iP_j,
 [X_0,P_i]=-iP_0P_i $. In this case, ${\cal R}_{10}\equiv (X_\mu,L_{\mu
\nu})$ constitute a Lie algebra or a maximal set of Killing vector fields 
realized on a pseudosphere in five-dimensional Minkowski-space; $ -{\xi_0}^2 
+{\xi_1}^2+{\xi_2}^2+{\xi_3}^2+{\eta^2}=constant. $  In what follows, 
let us call Lie algebra $\cal R$ constituted by Killing vector fields 
in general the space-time algebra of Snyder-Yang's type.

Among ${\cal R}_{15}$ in Yang's model, \eqQ, which will be preferably 
adopted 
in what follows, one finds that the time-like operators $ X_0, P_0,$ 
and 
$ M_i$  have continuous eigenvalues, contrary to those of the space-like 
operators $ X_i,P_i, L_{ij}$ and $N$, which have discrete eigenvalues. 
This 
statement is easily confirmed from the following fact that, for instance,
 $X_i$ and $X_0$ are rewritten as 
$$
\eqalign{
          & X_i= {1 \over i} \partder{}{\alpha_i},  \cr
          & X_0= {1 \over i} \partder{}{\alpha_0}
}
\eqn\eqR
$$
under fixed $k_i(=\sqrt{{\eta}^2 +{\xi_i}^2})$ and $ k_0(=\sqrt{{\eta}^2-
{\xi_0}^2})$, respectively, where $\xi_i =  \break k_i \sin {\alpha}_i,
 \eta = k_i \cos {\alpha_i}$ and $ \xi_0=k_0 \sinh {\alpha}_0$, $\eta 
=k_0 \cosh {\alpha}_0$. Eigenfunctions of $X_i$ and $X_0$ are given by 
$ \exp {(in_i \alpha_i)}$ and $ \exp {(i t\alpha_0) }$ with eigenvalues 
of integer $n_i$ and continuous time {\it t}, respectively. 

By using the above result, let us here consider a representation basis 
of the space-time algebra $\cal R$, i.e., quasi-local space-time bases,
 $Q_n$ :

$$
       Q_n = \prod_i \lbrace \exp{in_i \alpha_i}\rbrace \lbrace \exp{i
       t\alpha_0}          \rbrace f_n(-{\xi_0}^2+{\xi_1}^2+{\xi_2}^2 
       +{\xi_3}^2+{\eta}^2,\tau),
\eqn\eqS
$$ 
where $i$ runs from 1 to 3. At this point, one finds that each exponential 
factor on the right-hand side is, respectively, eigenstate of $i$-th 
space coordinate $X_i$ with discrete eigenvalue $ n_i$ and $X_0$ with 
continuous eigenvalue $t$, but basis function $Q_n$ itself being a product 
of these factors can no longer be eigenfunction of neither of $X_i's$ 
and $X_0$ on account of their noncommutativity, while the last factor 
$f_n$ being ortho-normal function of $(-{\xi_0}^2+{\xi_1}^2+{\xi_2}^2 
+{\xi_3}^2+{\eta}^2)$ and $\tau$ is clearly eigenfunction of each $X_i$ 
and $X_0$ with a vanishing value.

Furthermore, one sees the following basic relations
$$
    X_i Q_n  = n_i Q_n  + \sum_{j \neq i} {n_j\xi_j \xi_i \over 
                  ({\eta}^2+{\xi_j}^2)}  Q_n  +{t\xi_0 \xi_i \over (\eta^2
                  -\xi_0^2)} Q_n              
\eqn\eqW
$$
and
$$    
    X_0 Q_n = tQ_n +\sum_i {n_i\xi_i \xi_0 \over 
                  ({\eta}^2+{\xi_i}^2)}. 
\eqn\eqX
$$

From \eqW\ and \eqX, one obtains the following important result that 
 the expectation values of each $X_i$ and $X_0$ remain $n_i$ and $t$;
$$
         {\bar X_i} \equiv \langle  Q_n | X_i | Q_n \rangle  = n_i
\eqn\eqY
$$
with fluctuation ${\Delta X_i}^{(n)}$
$$
\eqalign
{      ({\Delta X_i}^{(n)})^2 & \equiv \langle Q_n | (X_i - n_i )^2 |Q_n 
\rangle  \cr
        &= \sum_{j \neq i} {n_j}^2 \langle Q_n|{{\xi_j}^2 {\xi_i}^2  
        \over 
               (\eta^2+\xi_j^2)^2}|Q_n\rangle  
          + t^2\langle Q_n|{{\xi_0}^2 {\xi_i}^2 \over (\eta^2-{\xi_0}^
          2)^2}|Q_n\rangle ,
}
\eqn\eqZ
$$
and 
$$
          {\bar X_0} \equiv \langle Q_n|X_0|Q_n\rangle  =t
\eqn\eqAA
$$
with fluctuation ${\Delta X_0}^{(n)}$
$$
\eqalign
{      ({\Delta X_0}^{(n)})^2 & \equiv \langle Q_n | (X_0 - t)^2 |Q_n 
\rangle  \cr
          &=\sum_i {n_i}^2 \langle Q_n| {{\xi_i}^2 {\xi_0}^2 \over 
          (\eta^2+\xi_i^2)^2            }|Q_n\rangle .
}
\eqn\eqAB
$$

In the preceding argument, one can easily generalize the spatial dimension 
3, tacitly assumed in the above arguments, to the arbitrary dimension 
by extending the dimension of Snyder-Yang's parameter space as $(\xi_0,
\xi_1,\xi_2,\cdots,\xi_p;\eta,\tau)$.
 
In what follows, we call the representation space spanned by the basis 
vectors
$|Q_n\rangle 's$ discussed in this section, Hilbert space I, in contrast 
to Hilbert Space II which will be introduced in section 4 in the usual 
quantum field-theoretical sense. It should be noted that the present 
quasi-local representation functions $Q_n$ in Hilbert space I possibly 
describe a four-dimensional excitation state quasi-localized in space 
and time directions, and enable us to find correspondence to a conventional 
local field, as will be discussed in the last section. 

\chapter{Matrix Representation on the light cone bases and Nonlocal Field}

As was stated in Introduction, nonlocal field $U$  also is defined as 
an operator on Hilbert space I on an equal footing with  $\cal R$ 
discussed in the preceding section. In this connection, it is quite 
important to notice that the representation basis functions such as $Q_n$'s,
 on which noncommutative algebra $\cal R$ and nonlocal field $U$ work 
as operators, can never be defined on a definite time value in general. 
One therefore encounters the entirely different situation from the familiar 
quantum mechanics, where any complete set of representation bases is 
defined on a fixed time.

At this point, it is quite instructive to remember that Dirac\Ref\D\dirac 
early pointed out the importance of a role of light cone time in terms 
of ``front'' in his systematic consideration on ``Forms of Relativistic 
Dynamics''. More recently, in the study of string 
\def\dirac{P.~A.~M.~Dirac, Rev.~Mod.~Phys.~{\bf 21}(1949), 392.}
field theories, light cone variables $X^\pm $ play an important role 
to describe the string as an extended object in space and time. Indeed,
 Kaku and Kikkawa \ref\kk showed in their first formulation of quantum 
field theory of
\def\kk{M.~Kaku and K.~Kikkawa, Phys.~Rev.~{\bf D 10}(1974), 1110.}
 string, that string fields and their interactions can be described on 
 a definite light cone time $X^+= \tau$ under the so-called light cone 
 gauge fixing. Furthermore, it is interesting to note that the recent 
 matrix model\refmark{\Bf} or quantum mechanics of D particles as 
 fundamental constituents of superstrings is also successfully formulated 
 in the ``infinite momentum frame." In accordance with these thoughts, 
 it seems promising to conjecture that any nonlocal field which is nonlocal 
 in space-time in general, becomes local with respect to time in the 
 infinite momentum frame so as to allow to describe its time development. 

This conjecture comes also from a simple consideration of special theory 
of relativity that any region in space-time in the usual Lorentz frame 
with a certain finite time width or time uncertainty ($\Delta X_0 
=\epsilon$)  may be seen in the infinite momentum frame, as a region 
with a vanishingly small width of light cone time, $\Delta X^+ =0 $. 
This fact strongly supports the view that any nonlocal field $U$ becomes 
to have a definite light cone time in the infinite momentum frame. 

In what follows, let us consider this possibility in the present 
noncommutative version of space-time, by assuming the following commutativity 
in general\foot{It should be noted that any light cone time $X^+$ defined 
in each Lorentz frame boosted in a certain spatial direction is proportional 
to time in the infinite momentum frame realized in the same direction.}
$$
       [X^+, U] = 0,       
\eqn\eqAG
$$
where $X^\pm \equiv {1 \over \sqrt 2} (X_0 \pm X_3)$ in the present 
four-dimensional space-time. 

According to this assumption, let us now take the following {\it light 
cone representation} bases $F_n(t)'s$, in which both $X^+$ and $U$ are 
diagonal, with eigenvalues of continuous $t$ and $u_n(t)$, respectively,
 that is, 
$$
         X^+ F_n(t) (\equiv i(\xi^+ \partder{}\eta + \eta \partder{}   
                       {\xi^-} ) F_n(\xi_i, \xi^\pm, \eta,\tau ;t)) = 
         t F_n(t)
\eqn\eqAH
$$
and 
$$
          U = \sum_{n,t} |F_n(t)\rangle  u_n(t) \langle  F_n(t)|,
\eqn\eqAI
$$
where $\xi^\pm \equiv {1 \over \sqrt 2}(\xi_0 \pm \xi_3)$ and $|F_n(t) 
\rangle$ 
being ket vector corresponding to the basis function $ F_n(t) \equiv 
F_n(\xi_i, \xi^\pm,\eta ,\tau ;t)$. In the above expression and hereafter,
 we tacitly use such a notation $\sum_t$ even for continuous parameter 
$t$ in place of $\int dt.$  

By remarking the expression $X^\pm = i(\xi^\pm \partder{}\eta + \eta 
\partder{} {\xi^\mp})$ and   
$$
         X^\pm \exp {({-it\eta \over \xi^\pm})} = t \exp({-it\eta \over 
         \xi^\pm}),
\eqn\eqAJ
$$
 one finds $F_n(t)$ 
 satisfying \eqAH\ is actually given by
$$
        F_n(t) = \exp ({-it\eta \over \xi^+}) g_n(\xi_i,\xi^+,\tau ;t)
\eqn\eqAK
$$
with suitable ortho-normalized functions ${g_n}$. 

At this point, it should be noted that any space-time operator $R$ belonging 
to  $\cal R$ other than $X^+$, must be represented in the following 
non-diagonal form in general 
$$
    R=\sum_{mn,tt'} |F_m(t)\rangle \langle  F_m(t)| R | F_n(t')\rangle 
     \langle  F_n(t')|.
\eqn\eqAL
$$ 

\chapter{ Second Quantization and Hamiltonian Formalism in Matrix Model 
of D-particles }

Now let us turn our attention to the fundamental problem how to determine 
the time development of nonlocal field $U$, or the time development of 
$u_n(t)$ given in \eqAI. According to the ordinary idea of quantum mechanics,
 we expect this to be done by means of the Hamiltonian formalism. Indeed 
we find an important clue for this in matrix model\refmark{\Bf} of 
D-particles formulated in the infinite momentum frame or light cone time 
as mentioned in the preceding section. Let us call our attention to the 
Hamiltonian presented in the latter theory: 
$$
    H = C \tr{\lbrace  { P_i P_i \over 2} + {1 \over 4} {[X_i, X_j]}^2 
    \rbrace},\eqn\eqAN
$$
where we neglected the terms of supersymmetric partner. Here, $X_i$ and 
$P_i$ 
$(i=1,2,3, \cdots, 9)$, on which tr works, are $N\times N$ matrices and 
interpreted to express transverse components of position coordinates 
and momenta of $N$ D-particles in the infinite momentum frame realized 
in the 11-th spatial direction in 11 dimensional space-time. Especially 
the diagonal matrix elements $\langle n|X_i|n\rangle  (n=1,2,\cdots,N)$ 
are understood as denoting literally {\it position coordinates}
\refmark{\winum} of the $n$-th D-particle. 

At first sight, these $N\times N$ matrices $X_i$ and $P_i$ noncommutative 
in general may be identifiable in the $N$ infinite limit with our space-time 
quantities $X_i$ and $P_i$ belonging to space-time algebra ${\cal R}$ 
which are also expressed in noncommutative matrix form, $ \langle F_m(t)| 
R |F_n(t')\rangle $ as seen in \eqAL. However, it should be emphasized 
that the former $N\times N$ matrices are interpreted to express position 
coordinates and momenta of $N$ D-particles, contrary to the latter ones 
which are indeed concerned with space-time quantities $\cal R$  {\it 
irrelevant} to individual objects involved in space-time.  One remembers 
a similar situation in the familiar second quantization procedure in 
quantum field theory, where a whole set of individual {\it position 
coordinates} of identical particles finally turns into {\it spatial 
coordinates} as arguments of quantized field operators.       

Bearing this point in mind, let us first try to carry out second 
quantization of identical D-particle system by introducing the 
second-quantized 
D-particle field, $\bf D,$  on the light cone basis according to \eqAI:
$$
          {\bf D} = \sum_{n,t} |F_n(t)\rangle  {\bf d}_n(t)  \langle  
          F_n(t)|,
\eqn\eqAO
$$
where light cone basis function $F_n(t)$ is defined by \eqAK, simply 
extending the space-time dimension from 4 to 11. ${\bf d}_n(t)$ and its 
adjoint ${{\bf d}_n(t)^\dagger}$ denote second-quantized operators 
satisfying the following commutation relations:
$$
            [{\bf d}_m(t), {{\bf d}_n(t)}^\dagger] =  
            \delta_{mn}
\eqn\eqAP
$$
and thus
$$
             [ \bf D, {\bf D}^\dagger ] = {\bf 1}.
\eqn\eqAQ
$$
Here one notices that D-particles are assumed to be identical particles 
obeying Bose-Einstein  statistics, and that ${\bf d}_n $ and its adjoint 
${{\bf d}_n}^\dagger $ are, respectively, annihilation and creation 
operators of 
excitation mode, $ |F_n \rangle $  of D-field. In what follows, let us 
call the Hilbert space, on which the above commutation relations are 
realized, Hilbert space II by distinguishing it from  Hilbert space I 
spanned by $|F_n(t)\rangle 's$ or $|Q_n\rangle 's$.  According to this 
definition, one finds that the second quantized D-field ${\bf D}$ is 
an operator both in Hilbert space I and II.

Now let us try to construct second-quantized Hamiltonian on a definite 
light cone time $t$ in terms of the $\bf D$ field defined in Hilbert 
space I, II and  space-time quantities, $X_i$ and $P_i$ defined in Hilbert 
space I, in close contact with Hamiltonian $H$ in \eqAN. At this point,
 let us regard the second term \foot{One can also regard this term as 
a kind of self-energy, which is neglected in the present paper.}on the 
right-hand side of \eqAN\  as the so-called two-particle term which 
expresses the interactions between two D-particles, in contrast to the 
first term expressing one-particle term concerning $N$ individual 
D-particles.  This consideration naturally leads us to the following 
form of second-quantized Hamiltonian ${\bf H}(t)$ defined on light cone 
time $t$ as an operator in Hilbert space II, which corresponds to $H$ 
in \eqAN\ of quantum mechanical many- D-particle system,
$$
\eqalign{
     {\bf H}(t) &={C \over 2} \Tr_t {\lbrace {\hskip 2 pt}{\bf D}^\dagger 
     P_iP_i                   {\bf D} \rbrace} 
            +{C \over 4} \Tr_t{\lbrace {\hskip 2 pt}{\bf D}^\dagger [X_i,
            X_j] 
             {\bf D} {\hskip 2 pt}{\bf D}^\dagger [X_i,X_j] {\bf D}\rbrace} 
             \cr
             &={C \over 2}\sum_n \langle  F_n(t) | P_iP_i | F_n(t)\rangle  
                {{\bf d}_n}(t)^\dagger {\bf d}_n(t) \cr
              &+{C \over 4}\sum_{mn} \langle F_n(t) | [X_i, X_j] | 
              F_m(t)\rangle              \langle  F_m(t)| [X_i,X_j] | 
              F_n(t)\rangle    \cr
              &\qquad \times {\bf d}_n(t)^\dagger {\bf d}_m(t) 
              {{\bf d}_m}(t)^\dagger {\bf d}_n(t),
}
\eqn\eqAR
$$
where $\Tr_t$ means trace concerned with sub-space of Hilbert space I 
spanned by basis vectors $| F_n(t)\rangle 's$ with fixed time $t$. In 
the above derivation of the second expression in \eqAR, it is crucial 
that $L_{ij} \equiv i[X_i, X_j]$ defined in \eqF\ is commutative with 
$X^+$, in restricting intermediate states to be on fixed light cone time 
$t$.

In order to examine directly the relation between $H$ in \eqAN, i.e., 
quantum-mechanical $N$-body Hamiltonian of D-particles in matrix model 
and the present quantum field-theoretic one ${\bf H}(t)$ given in \eqAR,
 let us first define quantum  field-theoretic $N$-body state at $t=0$ 
in terms of ${\bf d}_n (\equiv {\bf d}_n(0))$ as 
$$
    | N_1 N_2 \ldots ;N \rangle \rangle  \equiv A {{{\bf d}_1}^\dagger
    }^{N_1} 
        {{{\bf d}_2}^\dagger}^{N_2}{\ldots}  |0 \rangle \rangle , 
\eqn\eqAS
$$
with $N_1+N_2+ \ldots=N$, where $N_n$ denotes eigenvalue of number operator 
${\bf N}_n \equiv {{\bf d}_n}^\dagger {\bf d}_n $, A is a normalization 
constant and  $|0 \rangle \rangle  $ is the quantum field theoretic vacuum 
state of $D$-particle, that is, the ground state in Hilbert space II. 
Now it is easy to calculate the expectation value of ${\bf H}(0)$ with 
respect to this $N$-body state with the result: 
$$
\eqalign
{
 & \langle \langle N_1 N_2 \ldots ;N | {\bf H}| N_1 N_2 \ldots;N \rangle 
 \rangle                                                               
             \cr
        &= {C \over 2} \sum_n N_n \langle F_n |{ P_i P_i \over 2}|F_n\rangle 
                    +{C \over 4} \sum_{mn} N_m N_n \langle F_m | [X_i, 
        X_j] | F_n\rangle  
                \langle F_n | [X_i,X_j] |F_m\rangle 
}
\eqn\eqAT
$$
with $|F_n\rangle $ being $ |F_n(t=0)\rangle $, where we omitted the 
self-contraction term 
arising in the second term in the last expression. 

At this point, let us set all of $N_n$  to be equal to unity, then \eqAT\ 
becomes$$
      C \sum_n \langle  F_n | \lbrace {P_iP_i \over 2} + {1 \over 4}{[X_i,
      X_j]}^2 
       \rbrace | F_n \rangle .
\eqn\eqAU
$$

One should remark here that this expression formally coincides with the 
expression of $H$ in \eqAN, that is, the quantum mechanical Hamiltonian 
of $N$-body D-particle system with infinite $N$ limit, if one identifies 
the infinite dimensional matrix elements  $\langle m| R |n\rangle $ in 
 matrix model, where $m$($n$) is interpreted to mean the $m$($n$)-th 
$D$-particle, with the corresponding quantities $\langle F_m| R |F_n\rangle 
$, which are defined in Hilbert space I and interpreted as concerned 
with $m$($n$)-th {\it excitation mode} of D-field. This result strongly 
supports our choice of quantum field theoretic Hamiltonian of D-particles 
presented in \eqAR, if one takes into account that quantum mechanical 
$N$ D-particles are {\it identical particles} indistinguishable from 
each other, by virtue of the Weyl group of $U(N)$. \refmark{\winum}

Now we are in a position to consider the time development of D field, 
that is, Heisenberg's equation of motion. By noticing again $L_{ij} =i 
[X_i,X_j]$ defined in \eqF, it can be written down as follows
$$
\eqalign
{
  i{\hbar}{d \over dt} {\bf d}_n(t) &= [{\bf d}_n(t), {\bf H}(t)]  \cr 
     &={C \over 2}\langle F_n(t) | P_iP_i | F_n(t) \rangle  {\bf d}_n(t) 
     \cr
  &-{C \over 2} \sum_m \langle F_n (t)| L_{ij} |F_m(t) \rangle  \langle 
  F_m(t) | L_{ij} |F_n(t)\rangle   \cr     
    & \qquad \times  {\bf d}_m(t) {{\bf d}_m(t)}^\dagger {\bf d}_n(t), 
}
\eqn\eqAV
$$
where we omitted some self-contraction term. 

\chapter{Lorentz-invariant Action of {\bf D} field and Hamilton's Principle}

In this section, we propose Hamilton's principle of Lorentz-invariant 
action of {\bf D} field and investigate what conditions or approximations 
are needed to reproduce from it Heisenberg's equation of motion of {\bf 
D} field \eqAV\ given in non-covariant light cone frame. We imagine, 
from the explicit form of second-quantized Hamiltonian operator \eqAR, 
that there exists Hamilton's variation principle based on the Lorentz 
invariant action 
$I$: 
$$
            \delta I=0
\eqn\eqAW
$$
with
$$
    I= {1 \over 2}\lbrace \Tr({\bf D}^\dagger P_\mu {\bf D} P_\mu) 
       + \Tr(P_\mu {\bf D}^\dagger P_\mu {\bf D}) \rbrace 
        - \kappa \Tr{({\bf D}^\dagger L_{\mu \nu} {\bf D}{\bf D}^\dagger 
         L_{\mu \nu} {\bf D})},
\eqn\eqAX
$$
where Tr means trace with respect to Hilbert space I.  One sees immediately 
that $I$ possesses a wider class of invariances under transformations 
on Hilbert space I which include  Lorentz transformation generated by 
$\exp (i\epsilon_{\mu \nu}L_{\mu \nu})$ with infinitesimal parameters 
$\epsilon_{\mu \nu} $ and $L_{\mu \nu} \equiv (L_{ij}, M_i)$ given in 
section 2.

Now the variation with respect to ${\bf D}^\dagger$ gives, by omitting 
some self-contraction term,
$$
   {\delta}_{{\bf D}^\dagger} I \equiv \Tr [(\delta {\bf D}^\dagger)(P_\mu 
    {\bf D} P_\mu - 2 \kappa L_{\mu \nu} {\bf D} {\bf D}^\dagger L_{\mu 
   \nu} 
     {\bf D})]=0.
\eqn\eqAY
$$
At this point, in order to make Euler-Lagrange's equation, which is derived 
from the above variation principle, to come near Heisenberg's equation 
\eqAV\ given in light cone frame, let us rewrite the above equation in 
the following form, separating the so-called transverse part from 
longitudinal part in terms of the light cone variables:
$$
\eqalign{
  \Tr [\delta {\bf D}^\dagger
   & \lbrace  P_i {\bf D} P_i - \big( P^+ {\bf D} P^- +P^-{\bf D} P^+ 
   \big)  \cr
  &- 2\kappa L_{ij} {\bf D}{\bf D}^\dagger L_{ij} {\bf D}  \cr
  & +4\kappa \big( L_{i+} {\bf D}{\bf D}^\dagger L_{i-} {\bf D} 
                    + L_{i-} {\bf D}{\bf D}^\dagger L_{i+} {\bf D} \big) 
                     \cr    &-4\kappa L_{+-} {\bf D}{\bf D}^\dagger L_{-+} 
                    {\bf D} \rbrace] =0,
}
\eqn\eqBB 
$$
where indices $i$ or $j$ denote the transverse components running from 
1 to 9 and $\pm$ the longitudinal components ${1 \over \sqrt 2}(0 \pm 
11)$ in 11 dimensional space-time.

Next let us rewrite the above equation in terms of light cone bases $ 
|F_n\rangle 's$ , by assuming the existence of a particular solution 
{\bf D} of the non-covariant form \eqAO\ according to the discussion 
given in section 3. In fact, we assume here the solution to satisfy
$$
        [X^+,\bf D] =0 
\eqn\eqBC
$$
and further
$$
    P_i [P_i, {\bf D}]=0, \foot{If one starts from beginning by 
restricting oneself to Yang's space-time algebra where the relation 
$ [P_i^2, X^+ ]=0$ is algebraically guaranteed, one can choose 
the light cone bases $\ |F_n(t) >'s $ introduced in section 3 as 
common eigenstates of $X^+, P_i^2$ and ${\bf D}$ defined by \eqAO, 
so that there holds the relation $[P_i^2, {\bf D}] = 0$. Then one 
finds that, under the latter relation, the assumption 
$ P_i [P_i, {\bf D}] =0 $ introduced here turns out to be equivalent 
to $ [P_i, {\bf D}] P_i =0 $ or $ [P_i, [P_i, {\bf D}]]=0$.} 
\eqn\eqBD
$$
for a reason stated below, by taking into account the commutativity 
$[P_i,X^+]=0$ under Yang's space-time algebra, which we preferably adopt 
in what follows.

In this case, under the variation
$$
   \delta {\bf D}^\dagger \equiv \sum_{t,n} |F_n(t)\rangle  \delta {\bf 
   d}_n^\dagger(t)      \langle F_n(t)|,
\eqn\eqAZ 
$$ 
\eqBB\ leads us to the following Euler-Lagrange's equation:
$$
\eqalign{
   &\langle F_n(t)|P_i P_i|F_n(t)\rangle {\bf d}_n(t)   \cr
  &- \sum_{t',m} \lbrace \langle F_n(t)| P^+|F_m(t') \rangle \langle  
  F_m(t')|P^-| F_n(t) \rangle  \cr
  & \qquad + \langle F_n(t)|P^-|F_m(t')\rangle  \langle F_m(t') |P^+|F_n(t) 
  \rangle \rbrace {\bf d}_m(t')  \cr
   &-2\kappa \sum_m \langle F_n(t) |L_{ij}|F_m(t) \rangle \langle 
   F_m(t)|L_{ij} |F_n(t)\rangle  {\bf d}_m (t) {\bf d}_m^\dagger(t){\bf 
   d}_n(t)   \cr
   &+4\kappa \sum_{t',m}\lbrace \langle F_n(t) |L_{i+}|F_m(t') \rangle 
    \langle                              F_m(t') |L_{i-} |F_n(t)\rangle 
   \cr
   & \qquad \quad + \langle F_n(t) |L_{i-}|F_m(t') \rangle  \langle    
                             F_m(t') |L_{i+} |F_n(t)\rangle \rbrace
    {\bf d}_m (t') {\bf d }_m^\dagger(t'){\bf d}_n(t) \cr
   &-4\kappa \sum_{t',m} \langle F_n(t) |L_{+-}|F_m(t') \rangle  \langle 
                                F_m(t') |L_{-+} |F_n(t)\rangle
      {\bf d}_m (t') {\bf d }_m^\dagger(t'){\bf d}_n(t) =0.
}
\eqn\eqBE
$$

In the above expression, one should notice that the first term on the 
left-hand side is well written in the present form on account of \eqBD,
 so as to reproduce the corresponding term in \eqAV. With respect to 
the third term, it is important to note that $L_{ij}$ are commutative 
with $X^\pm$, $[L_{ij},X^\pm]=0,$ so the intermediate states are all 
retained on light cone time $t$. Secondly, since $L_{i+} (= i[X_i,X^+])$ 
appearing in the fourth term is also commutative with $X^+$, i.e., $[L_{i+},
X^+]=0,$ one sees again that $t'$ appearing in the intermediate states 
is set equal to $t$. Therefore, $\langle F_n(t)|L_{i+}|F_m(t)\rangle 
\equiv i\langle F_n(t) |[X_i,X^+]|F_m(t)\rangle =0,$ and thus the fourth 
term entirely disappears.

Before discussing the fifth term, let us consider the second term in 
\eqBE, from which we expect to extract the most interesting term including 
the time derivatives of ${\bf d}_n(t)$ corresponding to the left-hand 
side of \eqAV.  At this point, if one adopts Yang's space-time algebra 
as mentioned above, one finds that $ [P^+, X^+]=0,$  $[P^-, X^+]=iN$ 
and $[N,X^+] =iP^+.$  Consequently $ \langle F_n(t)| P^+|F_m(t') \rangle$ 
in the term can be written as 
$$
    \langle F_n(t)| P^+|F_m(t') \rangle \equiv P^+_{nm}(t,t')= P^+_{nm}(t) 
    \delta (t-t').
\eqn\eqBF
$$
On the other hand, $\langle F_m(t')| P^- |F_n(t)\rangle$ is also written 
down, through the repeated use of commutation relations mentioned above,
 in the following form$$
\eqalign{
    &\langle F_m(t')|P^-|F_n(t)\rangle  = {1 \over (t-t')} \langle F_m(t')| 
    [P^-,           X^+] | F_n(t)\rangle   \cr
    &={i \over (t-t')} \langle F_m(t')| N | F_n(t) \rangle ={-1 \over 
    (t-t')^2} 
      \langle F_m(t')| P^+ | F_n(t)\rangle. 
}
\eqn\BH 
$$
In this way one can rewrite the second term in \eqBE\ as 
$$
\eqalign{
   &- \sum_{t',m} \lbrace \langle F_n(t)| P^+|F_m(t') \rangle \langle  
   F_m(t')|P^-| F_n(t) \rangle  \cr
  & \qquad + \langle F_n(t)|P^-|F_m(t')\rangle  \langle F_m(t') |P^+|F_n(t) 
   \rangle \rbrace {\bf d}_m(t') \cr
    &=2 \sum_{t',m}{P^+_{nm}(t,t') P^+_{mn}(t',t) \over (t'-t)^2} {\bf 
    d}_m(t')  \cr
    &= -2 \sum_{t',m}{P^+_{nm}(t,t') \over (t'-t)} P^+_{mn}(t) \lbrace 
     {d \over dt'} \delta (t-t') \rbrace {\bf d}_m(t').
}
\eqn\eqBI
$$
One finds that the above singular expression includes the following term 
with 
first-order time derivative of ${\bf d}_n(t)$, in addition to second- 
and zeroth-order time-derivative terms, as
$$
     -2i \sum_m H_{nm}(t) {d \over dt} {\bf d}_m(t),
\eqn\eqBJ
$$
where 
$$
 H_{nm}(t) \equiv i P^+_{mn}(t) {\big(}{d \over dt'}P^+_{nm}(t,t'){\big)}
|_{t'=t}.
\eqn\eqBJA
$$

In combining the above results, \eqBE\ as a whole can be written in the 
following form
$$
\eqalign{
    & 2i \sum_m H_{nm}(t) {d \over dt} {\bf d}_m(t) \cr
    &=\langle F_n(t)|P_i P_i| F_n(t)\rangle  {\bf d}_n(t)   \cr
    &-2\kappa \sum_m \langle F_n(t) | L_{ij} |F_m(t) \rangle  \langle F_m(t)
      | L_{ij} |F_n(t) \rangle {\bf d}_m(t) {{\bf d}_m(t)}^\dagger {\bf 
    d}_n(t),}
\eqn\eqBK
$$
neglecting the second- and zeroth-order time derivative terms of ${\bf 
d}_n(t)$ mentioned above and one {\it anomalous} term which comes from 
the fifth term in \eqBE\ and will be discussed soon later. As a matter 
of fact, further neglecting the non-diagonal parts of $H_{nm}$, one finds 
that the above equation \eqBK\ reproduces ultimately \eqAV\ under the 
following correspondence
$$
            {{\hbar} \over C} \sim H_{nn}(t)               
\eqn\eqBL
$$
and $\kappa =1/2$. One might renormalize the $n$-dependent factor of  
$H_{nn}(t)$, if necessary, into ${\bf d}_n(t) $ and ${\bf d}^\dagger_n$,
 still keeping the commutation relations \eqAP\ and \eqAQ.

Finally let us explicitly calculate the {\it anomalous} term mentioned 
above, which comes from the fifth term on the left-hand side of \eqBE:
$$
   -4\kappa \sum_{t',m} \langle F_n(t) |L_{+-}|F_m(t') \rangle  \langle 
                                F_m(t') |L_{-+} |F_n(t)\rangle  
      {\bf d}_m (t') {\bf d }_m^\dagger(t'){\bf d}_n(t). 
\eqn\eqBM
$$
By remarking 
$$
          [L_{+-},X^\pm]=\mp iX^\pm,
\eqn\BN
$$
one finds
$$  \langle F_n(t) | L_{+-} |F_m(t')\rangle =it  {\delta(t-t')
     \over (t-t')}\delta_{mn}.
\eqn\eqBO
$$
Consequently, \eqBM\ can be written as
$$
   -4\kappa \sum_{t'} A(t,t') {\bf d}_n (t') {\bf d }_n^\dagger(t'){\bf 
   d}_n(t), 
\eqn\eqBP
$$
where anomalous factor $A(t,t')$ is defined by
$$
          A(t,t')\equiv {tt' \over (t-t')^2}\delta^2(t-t').
\eqn\eqBQ
$$
One finds that this is a kind of self-interactions of D particles with 
anomalous nonlocal time factor $A(t,t')$. 

At this point, it should be noticed that $L_{+-}$ appearing in $ \langle 
F_m(t') | L_{+-} |F_n(t)\rangle$ is the boost operator in the 11-th spatial 
direction $L_{0~11}$ and the light cone basis state $F_n(t)$ defined 
by \eqAH\ is understood as a limiting state attained at the maximal boost 
in the same direction. Consequently we conjecture that the origin of 
anomaly is in the overfull operation of boost on the light cone basis 
state. 

\chapter{Conclusions and Discussions} 

In this paper, we have proposed Hamilton's principle of relativistic 
action of matrix model, on the basis of a general framework describing 
nonlocal field associated with noncommutative space-time of Snyder-Yang's 
type. In section 4, we have found that both noncommutativities appearing 
in position coordinates of D particles in matrix model and in quantized 
space-time are eventually unified through second quantization of matrix 
model. In section 5, starting from the relativistic action principle 
of {\bf D} field, we have tried to reproduce Heisenberg's equation of 
motion \eqAV\ derived in section 4 in close contact with Hamiltonian 
in matrix model and clarified what conditions or approximations are needed 
for the purpose. Anomalous aspects encountered there, in self-interaction 
or self-mass terms , for instance, seem to be deeply concerned with the 
limiting character of light cone bases or the infinite momentum frame, 
while we do not enter into detailed discussions on these problems in 
the present paper.

 It should be noted here that space-time algebra of Yang' type has played 
 an important role in reproducing Heisenberg's equation \eqAV\ from 
 Hamilton's action principle mentioned above. In fact, in executing the 
 latter process in section 5, we have preferably adopted Yang's algebra,
  which has simple characteristic commutation relations of light cone 
 time operator $X^+$ with $P^{\pm}$, compared with Snyder's one.

In connection with the above argument, it is interesting to ask how the 
space-time algebra is to be chosen, or how the structure of Hilbert space 
I and II is 
possibly related, although in the present paper space-time algebra $\cal 
R$ is regarded as one given {\it kinematically} from the beginning, as 
in the form of Snyder's type, Yang's type and so on. There exists, however,
 another interesting possibility that space-time algebra $\cal R$ is 
Lie algebra composed of Killing vector fields realized in the parameter 
space of Snyder-Yang's type, whose metric structure is determined {\it 
dynamically} in connection with dynamics of matter field in accordance 
with the thought of general theory of relativity. 

This possibility may be actually formulated in applying Hamilton's principle 
to action $I$, by taking variations with respect to matrices of space-time 
quantities connected with Hilbert space I, in addition to the variation 
of ${\bf D}$ \eqAZ\ connected with Hilbert space II. This approach seems 
to be attractive, if we consider that matrix model includes gravitation. 
It may add a new light to the realization of noncommutative version of 
general theory of relativity or extended Kaluza-Klein theory,\ref\stanaka 
which possibly governs the world of Planck length.     

Finally let us comment on the correspondence between the present nonlocal 
field theory based on quantized space-time and a conventional local field 
theory. As was mentioned at the end in section 2, the correspondence 
may be well given by means of quasi-local representation bases $|Q_n\rangle 
's$, where quasi-local field  ${\bf U}$ may be expressed as $ {\bf U}=\sum_n 
|Q_n\rangle  {\bf u}_n \langle Q_n|$ analogously to \eqAO. One can imagine 
that ${\bf u}_n ({{\bf u}_n}^\dagger)$ denotes annihilation (creation) 
operator of {\it excitation} state $|Q_n\rangle $ within quasi-local 
four-dimensional region ( or ``elementary domain" according to Yukawa\ref\el)
and has correspondence to quantized local field ${\bf U}(x_\mu)$ 
with $x_\mu$ identified with expectation values of $X_\mu$ with respect 
to $|Q_n\rangle $ given in section 2. In the present paper, we have entirely 
omitted, for brevity of simplicity in expression, the explicit use and 
discussions of physical dimensions, for instance, both of $X_\mu$ and 
$P_\mu$ being defined in ${\eqB \sim \eqE}$ as no-dimensional ones. This 
important problem, however, must be left to future discussions, which 
might clarify, with the aid of \eqBL, the internal relation between Planck 
constant $\hbar$, Planck length and so on.  %
\def\el{H.~Yukawa, Suppl. Prog.~Theor.~Phys.~{\bf 37~\&~38}~(1966), 512; 
\nextline Y.~Katayama and H.~Yukawa, {\it ibid.}~{\bf 41}~(1968), 1;  
\nextline Y.~Katayama, I.~Umemura and H.~Yukawa, {\it ibid.}~{\bf 41}~(1968),
 22.}
\def\stanaka{S.~Tanaka, Prog.~Theor.~Phys. Suppl.~{\bf 67}(1979), 282.}

\ack{The author would like to thank S.~YAHIKOZAWA for the kind information 
and discussions on the recent development in matrix model and noncommutative 
geometry. He is deeply indebted to H.~AOYAMA for careful reading of the 
manuscript and giving valuable comments. Prof. S.~KAMEFUCHI kindly noticed 
the importance of the work of Dirac\refmark{\D} with respect to the present 
approach. Finally the author would like to thank Prof. S.~ISHIDA for 
constant encouragement of the study on space-time description of elementary 
particles.}
\endpage
\refout
\end